\documentclass{aa}
\usepackage{graphicx}
\usepackage{natbib}
\bibpunct{(}{)}{;}{a}{}{,}

\newcommand{\Msun}{M$_{\odot}$}

\newcommand{\mic}{$\mu$m}

\newcommand{\simless}{\mathbin{\lower 3pt\hbox
      {$\rlap{\raise 5pt\hbox{$\char'074$}}\mathchar"7218$}}} 

\begin{document}
   \title{Mid-IR spectroscopy of T Tauri stars in Chamealeon I: evidence for processed dust at the earliest stages}

   \subtitle{}

   \author{G. Meeus
          \inst{1}
          \and
          M. Sterzik\inst{2}
          \and
          J. Bouwman\inst{3}
          \and A. Natta\inst{4}\fnmsep\thanks{Based on observations with 
         TIMMI2 on the 3.6m at the European Southern Observatory, Chile; and on 
         observations with ISO, an ESA project with instruments funded by ESA 
         Member States (especially the PI countries: France, Germany, the 
         Netherlands and the United Kingdom) and with the participation of ISAS 
         and NASA.}
          }

   \offprints{G. Meeus}

   \institute{Astrophysikalisches Institut Potsdam, An der Sternwarte 16, D-14482 Potsdam\\
              \email{gwendolyn@aip.de}
         \and
             European Southern Observatory, Casilla 19001, Santiago 19, Chile\\
              \email{msterzik@eso.org}
         \and 
             CEA, DSM, DAPNIA, Service d'Astrophysique, CE Saclay, 91191 Gif-sur-Yvette Cedex, France\\
             \email{bjeroen@discovery.saclay.cea.fr}
         \and
             Osservatorio Astrofisico di Arcetri, INAF, Largo E. Fermi 5, 50125 Firenze, Italy\\
             \email{natta@arcetri.astro.it}
             }

   \date{Received ...; accepted ...}

   \abstract{We present mid-IR spectroscopy of three T Tauri stars in the
young Chamealeon I dark cloud obtained with TIMMI2 on the ESO 3.6m telescope. 
In these three stars, the silicate emission band at 9.7~\mic \ is prominent. 
We model it with a mixture of amorphous olivine grains of different size, 
crystalline silicates and silica. The fractional mass of these various 
components change widely from star to star. While the spectrum of CR Cha is 
dominated by small amorphous silicates, in VW Cha (and in a lesser degree in 
Glass~I), there is clear evidence of a large amount of processed dust in the 
form of crystalline silicates and large amorphous grains. This is the first 
time that processed dust has been detetected in very young T Tauri stars 
($\sim$ 1 Myr).
  
   }
\titlerunning{Processed dust in young TTSs}
   \maketitle

%

\section{Introduction}

Most studies of the dust composition in young stellar objects have concentrated
on Herbig Ae/Be stars (HAEBEs), as they are brighter than solar-mass T Tauri 
stars (TTS). Silicate emission at 10 $\mu$m has been detected in several TTS 
\citep{cohen1985ApJ...294..345C,1998ApJ...502..871H}, but the signal-to-noise 
of those observations was too low for an investigation of the properties of the 
emitting silicates. More recently, \citet{natta2000ApJ...534..838N} have 
obtained low-resolution mid-infrared spectra of 9 TTS in Chamealeon, using 
PHOT-S on board of ISO. All 9 stars showed silicate emission at 10~$\mu$m, 
interpreted as due to a mixture of amorphous silicates of radius $\simless$1 
$\mu$m on the surface of circumstellar discs. In some cases, there was a hint 
that a crystalline silicate component could be present, but the poor quality of
most spectra and their limited  coverage of the red side of the feature 
($\lambda < 11.7~\mu$m) prevented any detailed analysis.

Claims of the possible presence of a crystalline silicate component at 11.3 
$\mu$m have been made for some old TTS 
\citep[see, for example,][]{weinberger2002ApJ...566..409W}.  
However, the first clear evidence of crystalline materials in a low-mass object
has been obtained only this year for the 10 Myr-old TTS Hen3-600A by 
\citet{honda2003} using the 8.2m SUBARU telescope.

We are interested in extending the analysis of the silicate content of discs
to a large sample of TTS, possibly younger than Hen3-600A. In this letter, we 
present mid-IR spectra of three TTS in the young Chamealeon I cloud, already 
observed by \citet{natta2000ApJ...534..838N}. We show that the profiles are 
different in the three objects with a varying ratio of amorphous versus 
crystalline materials. In one star (VW~Cha) the ratio of crystalline to 
amorphous silicates is about 20\%, comparable to the values in the most 
evolved HAEBE stars and in solar system comets.

\section{Observations and data reduction}

We used TIMMI2 at the ESO 3.6m telescope to obtain N-band spectra of Glass~I, 
CR Cha (also known as LkH$\alpha$ 332-20) and VW Cha during the nights 
16-18/5/2003. These are three well-studied TTS with similar spectral types, 
all lying in the Chamealeon I dark cloud. In Table~1, we list the parameters 
of our sample stars \citep[from][]{natta2000ApJ...534..838N}. 
The mid-infrared spectra were obtained with standard chopping/nodding
techniques along the slit with a respective throw of 10 arcsec.  On source 
integration times were 1920 sec for Glass~I, 
2640 sec for CR Cha, 
and a total of 7200 sec for VW Cha. 
All observations were performed at airmasses below 2, typically around 1.6. 
In both nights, HD~133774 (K5III, N1 = 14.11), HD~169916 (K1IIIb, N1 = 
38.71), and HD~175775 (G8/K0II-III, N1 = 24.04) served as both telluric and 
flux standard stars and were observed at similar airmasses as the 
science targets. These standards are primary ISO calibration standard stars, 
and are described in detail at 
http://www.ls.eso.org/lasilla/Telescopes/360cat/timmi/html/stand.html.
The SED models used are according to \citet{cohen1998AJ....115.2092C}.
The atmospheric ozone feature at 9.54~$\mu$m was used for wavelength
calibration. The cross-calibration of the telluric standards among each
other shows a smooth and reliable removal of the atmospheric
bands, without introducing additional artifacts in the useful wavelength
regime of 8-13~$\mu$m. Spectrophotometric calibration was achieved by
observing each target in the N1 filter (central wavelength $\approx 8.6~\mu$m),
and normalising the integrated flux in the corresponding spectral 
passband to the observed value. We measured fluxes of 4.6$\pm $0.4~Jy for 
Glass~I, 0.5$\pm$0.2~Jy for CR Cha and 0.8$\pm$0.15~Jy for VW Cha.

\begin{figure}
\resizebox{\hsize}{!}{{\rotatebox{0}{\includegraphics{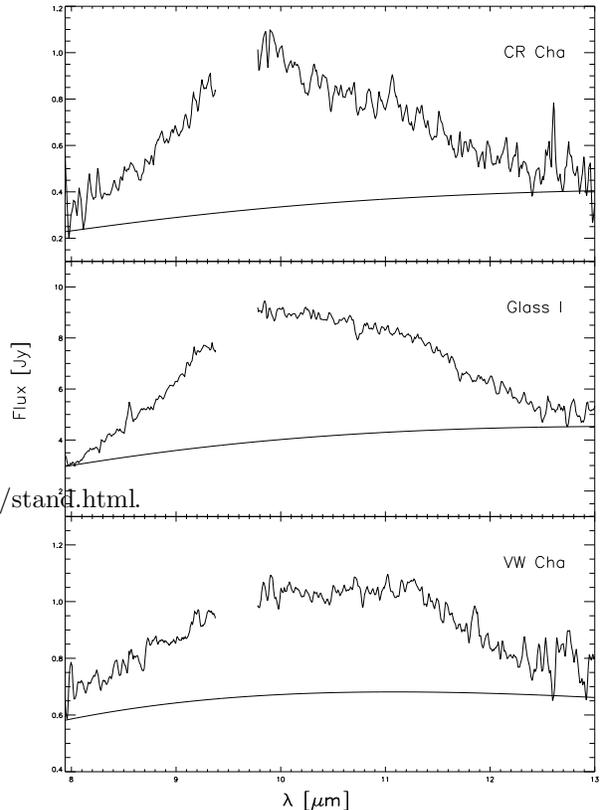}}}}
\caption{Observed TIMMI2 spectra of the three TTS, with a black body fit, 
representing the continuum.}
\label{spec}
\end{figure}

The spectra are shown in Fig.\ref{spec}. The profile of Glass~I, the only star
for which the ISO data have high quality, is very similar to that of 
\citet{natta2000ApJ...534..838N}. 

%
                                   
%

\section{Observed Sources}

A summary of the properties of the three observed stars is given in Table~1.
Unless otherwise stated, they are taken from \citet{natta2000ApJ...534..838N}.

\begin{description}

\item[\bf CR Cha] is a CTTS, for which no companions have been discovered 
\citep{takami2003A&A...397..675T}. 

\item[\bf Glass~I] is a binary (separation $2.''8$); the primary is a K4 WTTS 
  \citep{chelli1988A&A...207...46C}, the secondary an IR G-type object 
  \citep{covino1997A&AS..122...95C}. \citet{stanke2000} 
  attribute the 10~\mic \ emission to the secondary. Since the secondary is not
  visible in the optical, the age is estimated from the optical
  properties  of the primary,  assuming that they are coeval.

\item[\bf VW Cha]  is a $0.''7$ binary, with both members being CTTS; the IR 
  emission is attributed to the primary. Using 
  adaptive optics, \citet{brandeker2001} discovered a companion to the 
  secondary at a separation of $0.''1$. The age uncertainty for this object
  derives primarily in the uncertainty of the extinction towards VW Cha: 
  \citet{brandner1997} give A$_{\mathrm{V}} \sim$ 1.3 while 
  \citet{brandeker2001} state 3.0; the larger the adopted extinction, the 
  younger the derived age of the object. Most likely, the age of 
  the system lies between 2 and 0.4 Myr, values derived by these authors.
 
\end{description}

\begin{table}\label{tabel}
\begin{tabular}{lccccc}
\hline
\hline
Object & Spectral      &T$_{\mathrm{eff}}$ & L$_{*}$     & M$_{*}$ & Age  \\
       & Type          &(K)         &(L$_{\odot}$)& (\Msun) & (Myr) \\
       &               &            &             &         &       \\
\hline
CR Cha & K2            &4900        & 3.3     & 1.2       & 1.0  \\
Glass I& K4            &4600        & 1.6     & 0.9       & 1.0  \\
VW Cha & K5/K7         &4350        & 2.9     & 0.6       &1.0$^{+1}_{-0.6}$\\
\hline  
\end{tabular}    
\caption{Stellar parameters from \citet{natta2000ApJ...534..838N}, apart 
from the age of VW Cha, which is discussed in the text.} 
\end{table}     
%
 
\section{Modelling and results}

\begin{figure}
\resizebox{\hsize}{!}{{\rotatebox{0}{\includegraphics{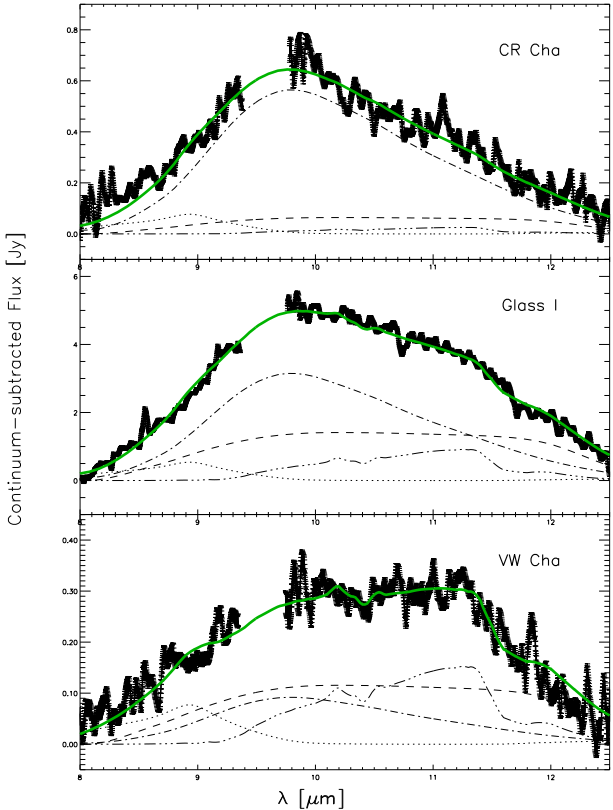}}}}
\caption{Fit to the continuum-subtracted spectra, with each separate component 
shown. We also show the formal error on the data, which has been determined from 
the background. The true errors are likely to be larger as the division by the 
standard star might produce artefacts from telluric features. Light grey: sum of 
all components, dash dot: small ($a$ = 0.1~\mic) amorphous silicate, dashed: 
large ($a$ = 2.0~\mic) amorphous silicate, dotted: silica; and dash triple dot: 
forsterite.}
\label{decompo}
\end{figure} 

To determine the composition of the circumstellar (CS) dust in our TTS, we
have adopted the procedure successfully used in interpreting the ISO spectra
of HAeBe stars by \citet{2001A&A...375..950B}. We first fitted a black body to
the continuum, as shown in Fig.\ref{spec}. The determination of the continuum
is difficult due to a lack of reliable measurements outside of the feature, and 
we estimate that it could be higher by 10\% for VW Cha, the most 
difficult case. This continuum uncertainty, however, has only a negligible 
effect on the results discussed in the following. We fitted the 
continuum-subtracted spectra with a linear combination of several dust species,
namely:

\begin{enumerate}
 \item{amorphous olivine ([Mg,Fe]$_{2}$SiO$_{4}$) with a size of 0.1 and 
 2.0~\mic, representing small and large grains,}
\item{crystalline silicates: magnesium forsterite (Mg$_{2}$SiO$_{4}$) 
   and pure magnesium enstatite (MgSiO$_{3}$)},
 \item{silica (SiO$_{2}$)}.
\end{enumerate}

\noindent
These species are commonly found in the dust discs of Herbig Ae/Be systems.
Details of the procedure and references to the adopted cross sections can be 
found in \citet{2001A&A...375..950B}. We did not include Polycyclic Aromatic 
Hydrocarbons (PAHs), since the PHOT-S spectra did not detect any feature 
between 6 and 8 $\mu$m, where the signal-to-noise was quite good 
\citep{natta2000ApJ...534..838N}. 

\begin{table}\label{table2}
\begin{center}
\begin{tabular}{cccccccc}
\hline
\hline
Object &BB &m$_{2.0}$/m$_{0.1}$&m$_{\mathrm{forst}}$/m$_{\mathrm{sil}}$&m$_{\mathrm{SiO_{2}}}$/m$_{\mathrm{sil}}$ \\
\hline
CR Cha & 350 K & 0.23              &0.02                 &0.04  \\
Glass I& 380 K & 0.92              &0.08                 &0.03  \\
VW Cha & 460 K & 2.57              &0.19                 &0.07  \\
\hline
\end{tabular}
\caption{Temperature of the black body (BB) representing the continuum and 
mass ratios: large to small amorphous silicates (m$_{2.0}$/m$_{0.1}$), 
crystalline to amorphous silicates (m$_{\mathrm{forst}}$/m$_{\mathrm{sil}}$) 
and silica to silicates (m$_{\mathrm{SiO_{2}}}$/m$_{\mathrm{sil}}$).}
\end{center}
\end{table}

Fig.~2 shows the fits to the 10~\mic \ spectra together with the separate 
contributions of each dust component; the derived mass ratios of the 
different species are listed in Table~2. In our final fit, we did not 
include enstatite, as it did not improve the fitting result. 
It is immediately clear that different species contribute in different amounts 
to the overall shapes.
The values of these ratios vary significantly from one star to the other. In 
particular, VW Cha has the largest fraction of large to small amorphous silicate
grains - 11 times that of CR Cha and 3 times that of Glass~I, and the largest 
amount of crystalline silicates and silica - a factor 10 and 2 more for 
crystalline silicates and a factor 4 and 2 more for silica than CR Cha and 
Glass~I, respectively. In contrast, the spectrum of CR Cha is dominated by 
small amorphous silicate grains. Glass~I is intermediate between the two. 
\citet{natta2000ApJ...534..838N} already noticed excess emission in 
their Phot-S spectrum of Glass~I at 8.5 and 11.3~\mic \ and attributed this
tentatively to silica and crystalline silicates. Our higher-quality spectrum
of Glass~I confirms this suggestion. 

\section{Discussion}

\begin{figure}
\resizebox{\hsize}{!}{{\rotatebox{0}{\includegraphics{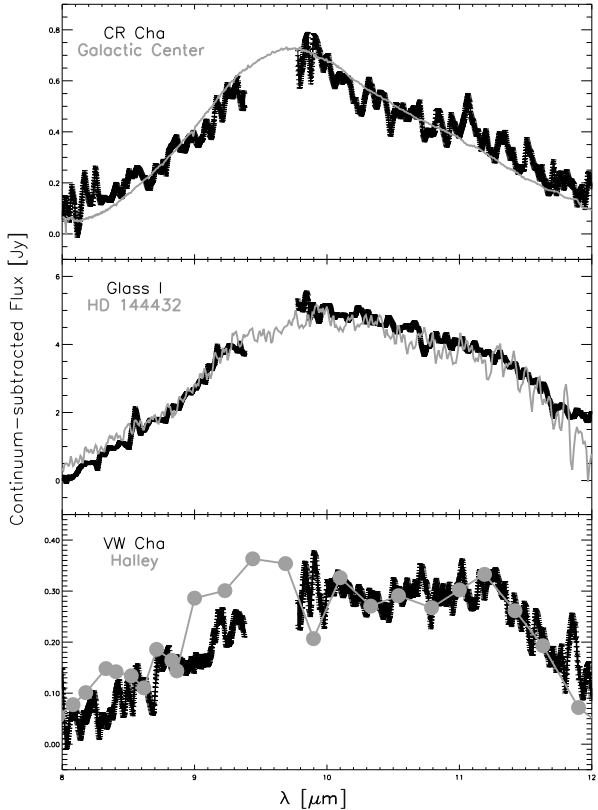}}}}
\caption{Comparison of the Cha TTS spectra with that of different objects. 
{\bf Top}: CR Cha with ISM silicate bands (inverted absorption 
band profile towards Sgr A*), which are dominated by amorphous silicates (at 
least 99 \%, Kemper, de Vriend \& Tielens, in prep.), {\bf Middle}: Glass~I with
the HAe star HD~144432, {\bf Bottom}: VW Cha with comet Halley 
\citep[data from][]{bregman1987A&A...187..616B}, which is a source of highly 
processed dust \citep{2001A&A...375..950B}. Halley's data point at 9.9~\mic \ 
has a problematic ozone subtraction, according to the authors. The sequence 
shows an increasing amount of processed dust.}
\label{haebe}
\end{figure} 

\begin{figure}
\resizebox{\hsize}{!}{{\rotatebox{0}{\includegraphics{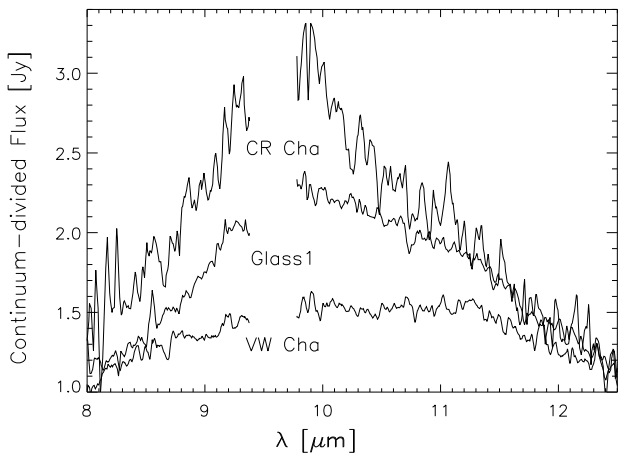}}}}
\caption{Continuum-divided spectra of the 3 objects. VW Cha clearly has the
weakest feature, CR Cha the strongest.}
\label{decompo}
\end{figure} 

We compared our TTS spectra with objects of different evolutionary status:
the ISM, a sample of 14 HAEBEs studied by \citet{meeus2001} and a few solar 
system comets. We found a good similarity between some objects 
and our TTS, as is shown in Fig.~3. These comparison objects were modelled by 
\citet{2001A&A...375..950B} who showed that they cover a broad range in the 
amount of processed dust, going from practically amorphous (the ISM) to highly 
processed (comet Halley). 

Even more interestingly, we find trends in our small sample of TTS similar to 
those that have already been identified in HAEBEs. The first of such trends is 
an {\em anti-correlation between silicate grain size and strength of the band}, 
interpreted as a dust evolutionary process (van Boekel et al. 2003). Fig.~4 
shows that this is also the case for the three TTS. VW Cha has the weakest 
feature and the smallest mass fraction of small silicate grains. CR Cha, on the 
other hand, has the strongest feature and the largest mass fraction of small 
grains. 

In a study of the 10~\mic \ feature, \citet{2001A&A...375..950B} derived a {\em 
correlation between the amount of silica and that of forsterite}, and suggested 
a common origin in the process of thermal annealing. This process is responsible
for the conversion of amorphous silicates into crystalline silicates, with 
silica as a by-product. We find a similar correlation in our data, as VW Cha has
both the largest amount of silica and of crystalline forsterite. The ubiquity of
the correlation is interesting, also in view of the well-known difficulty with 
the thermal annealing hypothesis, since the observed crystalline dust appears to
be colder than the minimum temperature required for such a process to be
efficient. Laboratory experiments suggest a temperature of 1000~K to anneal 
amorphous silicates into crystalline on a time-scale of a month, while a 
slightly lower temperature (875~K) would require more than 30 Myr 
\citep{hallenbeck2000}. This means that the dust, once crystallised, should be 
somehow transported to colder, more outward disc regions. Radial mixing by 
diffuse transport in CS discs has been proposed as a way to do it 
\citep{2002A&A...385..181W}.

We find no conclusive evidence for the presence of enstatite in our TTS spectra,
apparently contradictory to the results of \citet{honda2003} who found evidence 
for a substantial mass fraction of enstatite dust in the Hen3-600A system. 
However, among HAEBEs the presence of enstatite seems to be rare as well, as 
only one out of 14 studied objects (HD~179218) showed evidence for this dust 
species \citep{2001A&A...375..950B}. This is probably due to its formation
process: annealing experiments of silicate smokes \citep[e.g.][]{rietmeijer1988LPSC...19..513R,hallenbeck2000} show that the initially formed forsterite and 
silica can produce enstatite only on a considerably longer time scale. 
Only around objects which are much more luminous, such as evolved stars or 
HD~179218, or considerably older, such as Hen3-600A, the time-temperature 
conditions would be right to readily form large amounts of enstatite.

Our results indicate clearly that dust evolves in TTS in ways very
similar to those observed in HAEBE stars. The evolution does not seem
to depend on the luminosity, temperature or mass of the central star.
It is also difficult to reconcile our data with the idea that ``age"
is the only dominant factor.  Our observations clearly show that the
dusty disc of VW Cha is host of a lot of warm processed dust. CR Cha,
in contrast, has a similar age and spectral type but a lot of
unprocessed dust. All our three stars are about ten times younger than
Hen3-600A, the only other TTS where crystalline dust was firmly
detected. 


Interestingly, we note that VW Cha is a close binary system, in which the 
secondary is also a binary. The disc is circumprimary and is probably tidally 
truncated by the secondary binary, as it is not detected at longer ($\lambda 
\sim$ 60~\mic) wavelengths \citep{brandeker2001}. Glass~I, which has 
intermediate properties, is also a binary, but with a larger separation than VW 
Cha. {\em It is possible that interaction with close companions may speed up 
dust evolution} as it induces vertical mixing, stirring up the dust from the 
midplane. This causes a (continuous) replenishment of unprocessed dust into the
disc atmosphere where dust processing in the form of thermal annealing is 
expected to take place. Also, we would observe IR emission from larger grains as
grain growth occurs faster in denser regions, such as the midplane.
That dust processing in close binaries would occur faster is for the moment just
speculation, since a much larger sample of stars is obviously required 
in order to investigate its viability.

\section{Conclusions}

The 10 $\mu$m emission feature is clearly detected in three young TTS in 
Chamealeon I. The dust composition, obtained by fitting the shape of the
feature with a mixture of materials, derived from detailed studies of HAEBE 
stars, differs significantly in the three stars. In particular, the spectrum 
of CR Cha is dominated by small, amorphous silicates. The spectrum of VW Cha 
shows evidence of highly processed dust, as shown by the high mass fraction of 
large amorphous silicates and of crystalline ones. Glass~I seems to be 
intermediate between the two other stars.

This trend, as well as the correlation between the relative amount of small
grains and the intensity of the feature and that between the amount of silica 
and forsterite, are also found in HAEBE stars. Also, as in HAEBE stars, 
dust evolution does not seem to be uniquely controlled by the stellar age. Our 
Chamaeleon sample consists of three stars of practically the same spectral type
and age and widely different dust. Looking for possible clues to this unknown 
``trigger" of dust evolution, we noticed a correlation in our stars between 
dust processing and the presence of close companions. Although a sample of 
three stars is by far too small to provide any evidence, we suggest that this 
possibility should be explored further.


                               
\begin{acknowledgements}
GM and JB acknowledge financial support by the EC-RTN on ``The Formation and
Evolution of Young Stellar Clusters'' (RTN--1999--00436,
HPRN--CT--2000--00155). A.N. was partly supported by ASI grant ARS-1/R/073/01. 
This research has made use of NASA's Astrophysics Data System Bibliographic 
Services and the SIMBAD database, operated at CDS, Strasbourg, France.

\end{acknowledgements}

\bibliographystyle{/z/gwen/ARTICLES/A_and_A/bibtex/aa} 
\bibliography{/z/gwen/ARTICLES/A_and_A/bibtex/bibliofile} 

\end{document}